\documentclass[12pt]{article}
\usepackage[textures]{graphics}
\parskip=\bigskipamount
\newcommand{\beq}{\begin{equation}}
\newcommand{\eeq}{\end{equation}}
\newcommand{\beqa}{\begin{eqnarray}}
\newcommand{\eeqa}{\end{eqnarray}}
\newcommand{\ket} [1] {\vert #1 \rangle}
\newcommand{\bra} [1] {\langle #1 \vert}
\newcommand{\braket}[2]{\langle #1 | #2 \rangle}

\begin{document}
\title{\bf Derivation of an explicit expression for mutually unbiased bases in even and odd prime power dimensions.}
\date{}
\author{}
\maketitle
\vglue -1.8truecm
\centerline{\large Thomas Durt\footnote{TENA-TONA Free University of
Brussels, Pleinlaan 2, B-1050 Brussels, Belgium. email: thomdurt@vub.ac.be}}\bigskip\bigskip


{\it Abstract: Mutually unbiased bases generalize the $X$, $Y$ and $Z$ qubit bases. They possess numerous applications in 
quantum information science. 
It is well-known that in prime power dimensions $N=p^m$ (with $p$ prime and $m$ a positive integer) there exists a maximal set of $N+1$ mutually unbiased bases. 
 In the present paper, we derive an explicit expression for those bases, in terms of the (operations of the) associated finite field (Galois division ring) 
 of $N$ elements. This expression is shown to be equivalent to the expressions previously obtained by Ivanovic
  in odd prime dimensions (J. Phys. A, 14, 3241 (1981)), and Wootters and Fields (Ann. Phys. 191, 363 (1989)) 
  in odd prime power dimensions. In even prime power dimensions, 
  we derive a new explicit expression for the mutually unbiased bases. The new ingredients of our
   approach are, basically, the following: we provide  a simple expression of the generalised Pauli group in terms of the additive characters
    of the field and we derive an exact groupal composition law
     inside the elements of the commuting subsets of the generalised Pauli group, 
    renormalised by a well-chosen phase-factor. }   

\section*{Introduction}

Two orthonormal bases of a $N$ dimensional Hilbert space are said to be mutually 
  unbiased if whenever we choose one state in the first basis, and a second state in the second basis,
   the modulus squared of their in-product is equal to $1/N$.  It is well-known that, when the dimension of the Hilbert space is a prime power, there exists a 
   set of $N+1$ mutually unbiased bases.  This set is maximal because it is not possible to find 
   more than $N+1$ mutually unbiased bases in a $N$ dimensional Hilbert space \cite{Ivanovic,Wootters,india}. It is also a complete
    set because when we know all
    the probabilities of transition of a given quantum state towards the states of the bases of this set (there are $N^2-1$ of them), we can reconstruct all the coefficients of the density matrix
    that characterizes this state; in other words we can perform full tomography or complete
     quantum state determination \cite{Ivanovic,Wootters,Schwinger,Yu}. 
   A crucial element of the construction is the existence of a finite commutative division ring
    (or field\footnote{A field is a set with a multiplication and an addition operation which satisfy the usual rules, associativity and commutativity
    of both operations, the distributive law, existence of an additive identity 0 and a multiplicative identity 1, additive inverses, and multiplicative 
    inverses for every element, 0 excepted. })
   of $N$
   elements. As it is well known, finite fields with $N$ elements exist if and only if the
     dimension $N$ is a power of
     a prime, and a derivation of a set of mutually
      unbiased bases is already known in such cases. Note that nobody managed until now to generalize this construction in the absence of finite field so that 
   it is still an open question whether such sets exist when the 
   dimension is not a prime power \cite{Archer,Grassl}. In the present paper, we obtain, in a synthetic formulation, the expressions for the mutually unbiased bases 
   that were derived in the past. In odd prime power dimensions, we recover by a slightly different approach the expressions already obtained in the past by 
   Ivanovic \cite{Ivanovic} and in odd 
   prime power dimension $p^m$ by Wootters and Fields \cite{Wootters}). We provide a synthetic expression that is also valid in even prime powers dimensions ($2^m$). 
   The (discrete)¥ Heisenberg-Weyl group 
   \cite{india,Schwinger,weyl} (sometimes also called generalized Pauli group)¥, a finite group of unitary transformations, plays a central role in our
    approach.

    \section{Preliminary concepts}
  In what follows, we shall systematically assume that we work in a Hilbert space of prime power dimension $N=p^m$ with $p$ a prime number, and $m$ a positive integer¥¥. Then, 
  as is well known, it is possible to find a finite field with $N$ elements. We shall label these elements by an integer 
  number $i$, $0\leq i\leq N-1$), or, equivalently, by a $m$uple of integer numbers $(i_{0},i_{1},...,i_{m-1})¥$¥ 
  running from 0 to $p-1$ that we get from the $p$-ary expansion of $i$: $i=\sum_{k=0}^{m-1}i_{n}p^n$¥¥. 
  This field is characterized by two operations, a multiplication and an addition, that we shall denote 
 $\odot_G$ and $\oplus_G$ respectively. It is always possible to label the elements of the field in such a way that the addition is equivalent with the addition modulo $p$¥ componentwise. 
 As all the fields are equivalent, up to a relabelling, there is no strict obligation to do so, but it is more natural and convenient. This property is a direct consequence of the fact that for all the finite 
 fields the characteristics of the field,
   which is the smallest number of times that we must add the element 1 (neutral for the multiplication)¥with itself before we obtain 0 (neutral for the addition)¥, is always equal to a prime number
    ($p$ when $N=p^m$¥).
   The index $G$ refers to Evariste Galois and is introduced in order not to confuse
   these operations with the 
 usual (complex) multiplication and addition 
  for which no index is written.
  
Let us denote $\gamma_{G}$ the $p$th root of unity: $\gamma_{G} =e^{ i .2\pi/p}$. 
Exponentiating  $\gamma_{G}$ with elements $g$ of 
the field (with the usual rules for exponientation), we obtain complex phasors of the type 
$\gamma_{G}^{ g}$ ($0\leq g\leq N$). Such phasors can take $p$ different values. They can be
 considered as a $p$-uple 
generalisation of the (binary) parity operation $e^{ i .(2\pi/2).g}$ that corresponds to the qubit case in the sense
 that the phasor $\gamma_{G}^{ g}$ ($0\leq g\leq N$) only depends on the value of 
   the first component $g_{0}¥$ of the $p$-ary expansion of $g$ which is nothing else than the remainder of $g$ after division by $p$,
    when the division by $p$ is taken in the usual sense.
   
The following identity appears to play a fundamental role in our approach:   
   \beq\label{identi1}\sum_{j=0}^{N-1} \gamma_{G}^{ (j\odot_{G} i)}=N\delta_{i,0}\eeq 
Indeed, if $i=0$, then $\sum_{j=0}^{N-1} \gamma_{G}^{ (j\odot_{G} i)}=N.1=N$. Otherwise,
 $\sum_{j=0}^{N-1} \gamma_{G}^{ (j\odot_{G} i)}=\sum_{j'=0}^{N-1} \gamma_{G}^{ j'}$ in virtue of the inversibility of the multiplication. Now the exponentiation of gamma by elements of the field does only depend on the remainder after division by $p$, so that 
 $\sum_{j'=0}^{N-1} \gamma_{G}^{ j'}=¥p^{m-1}.\sum_{j'=0}^{p-1} \gamma_{G}^{ j'}=p^{m-1}.{(1-\gamma_{G}^m)\over (1-\gamma_{G})}=0.$
  
  In virtue of the fact that the addition is the addition modulo $p$, componentwise, we can derive the following identity which is also very useful:
 
 \beq\label{identi2}\gamma_G^{i}\cdot\gamma_G^{j}=\gamma _{G} ^{(i\oplus_G j)}\eeq
 Indeed, $\gamma_G^{i}\cdot\gamma_G^{j}=\gamma _{G} ^{(i+ j)}=\gamma _{G} ^{(i_{0}¥+ j_{0}¥)}=
 \gamma _{G} ^{(i\oplus_G j)_{0}¥}=\gamma _{G} ^{(i\oplus_G j)}$ (in the previous expression, we represented by the symbol 
 $x_{0}¥$ the remainder of $x$ after division
  by $p$, where $x$ is an element of the field, comprised between 0 and $p^m-1=N-1$,
   and the division by $p$ is taken in the usual sense.) This relation is well-known and
    expresses, in the language of mathematicians, that $p$th roots of unity are additive 
    characters of the Galois field \cite{Karpilovsky}.¥
  
  It is important to note, in order to avoid confusions, that different types of operations are present at this level: the internal field operations are labelled
   by the lower index $G$. 
  They must not be confused with the modulo $N$ operations. 
  In order to emphasise the difference between these operations, 
  we give in example the corresponding tables in the case $N=4=2^2$ in appendix. One can check that the field and modulo 4 multiplications 
  are distributive relatively to the associated addition, but that there are no dividers of 0, 0 excepted, only in the case of the field multiplication. 
  As a consequence, the field multiplication table, amputed from the first line and column exhibits an invertible  (group) structure.
  All operations are commutative as can be seen from the symmetry of the tables 1 to 4 under transposition. 
  
  Remark that if we express quartits as products of two qubits: 
$\ket{0}_{4}=\ket{0}_{2}\otimes\ket{0}_{2}$, $\ket{1}_{4}=\ket{0}_{2}\otimes
\ket{1}_{2}$, 
$\ket{2}_{4}=\ket{1}_{2}\otimes\ket{0}_{2}$, $\ket{3}_{4}=\ket{1}_{2}\otimes\ket{1}_{2}$. It is then easy to check the following property:
 If $\ket{i}_{4}=\ket{i_{1}}_{2}\otimes\ket{i_{2}}_{2}$, and $\ket{j}_{4}=
 \ket{j_{1}}_{2}\otimes\ket{j_{2}}_{2}$, 
 then $\ket{i\oplus_{G} j}_{4}=\ket{i_{1}\oplus_{mod 2}j_{1}}_{2}\otimes\ket{i_{2}\oplus_{mod 2}j_{2}}_{2}$. 
 This is an illustration of the fact that the field addition is equivalent with the addition modulo
  $p$ componentwise. 
   It is also worth noting that the property $\sum_{p=0}^{N-1} \gamma^{ (p\odot q)}=N\delta_{q,0}$ is true for the modulo $N$¥ multiplication as well, but $ \gamma$¥ must be taken to be equal to the $N$th root of unity in this case.
  In prime dimensions $\gamma _{G}$ is the $N$th root of unity and the Galois and modulo $N$¥ operations coincide. 
 In prime power but non-prime dimensions, this is no longer true. 

  \section{Construction of the dual basis}

Let us now consider the unitary transformations  $V^0_{l}$, that shift each label of the states of the computational basis
 ($\{\ket{0},\ket{1},...,\ket{i},...,\ket{N-1}\}$) by a distance $l$¥ 
($\ket{i} \to \ket{i\oplus_{G}l})$ (the reason for our choice of notation will be made obvious soon). The transformations $V^0_{l}$
 form a commutative 
group with $N$ elements that is isomorphic to the Galois addition. 
  Generalizing the procedure outlined in \cite{DurtNagler}, 
  we define the dual basis as follows: 
   \beqa \ket{\tilde j}=\frac{1}{\sqrt{N}}\sum_{k=0}^{N-1} 
   \gamma_{G}^{\ominus_{G}(k\odot_{G} j)}\nonumber \\
   \ket{k } \label{dual}\eeqa where the symbol  $\ominus_{G}$ represents the inverse of the Galois 
addition  $\oplus_{G}$.
It is easy to check that the dual states are invariant, up to a global phase, under 
the transformations $V^0_{l}$. Indeed, we have:
\beqa
V^0_{l}.\ket{\tilde  j}=\frac{1}{\sqrt{N}}\sum_{k=0}^{N-1}
 \gamma _{G} ^{\ominus_{G}(k\odot_{G} j) }\ket{k \oplus_{G} l}\\
=\frac{1}{\sqrt{N}}\sum_{k'=0}^{N-1}\gamma _{G} ^{(\ominus_{G}(k'\ominus_{G}l)\odot_{G} j) }
\ket{k' }
=\gamma_{G}^{(l\odot_{G} j) }\ket{\tilde  j}\eeqa  Obviously, the dual basis and the computational basis are mutually
 unbiased. When the dimension is prime ($N=p$), the dual basis is the discrete
  Fourier transform of the computational basis, when it is a power of 2, it is a Hadamard transform \cite{DurtNagler}. 
  
 Let us denote $V_0^{l}$ the unitary transformations that shift each label of the states 
 of the dual basis
 ($\{\ket{ \tilde 0},\ket{\tilde 1},...,\ket{\tilde i},...,\ket{\tilde N-\tilde 1}\}$) 
 by a distance $\ominus_{G}l$¥ 
($\ket{\tilde i} \to \ket{{\tilde i\ominus_{G}\tilde l}})$. The transformations $V^l_{0}$
 form a commutative 
group with $N$ elements that is isomorphic to the Galois addition. It is easy to check that these operators are diagonal in the computational basis:

 \beq \label{transla1}V^l_{0}=  \sum_{k=0}^{N-1} \ket{\tilde k\ominus_{G}\tilde l}\bra{ \tilde k}
 =  \sum_{k=0}^{N-1} \gamma_{G}^{(k\odot_{G} l)}\ket{ k}\bra{ k}.\eeq 
This is the dual counterpart of a similar expression for the shifts in the computational basis: 
  \beq \label{transla2} V^0_{l}=  \sum_{k=0}^{N-1} \ket{ k\oplus_{G} l}\bra{  k}
 =  \sum_{k=0}^{N-1} \gamma_{G}^{(k\odot_{G}l) }\ket{\tilde k}\bra{\tilde k}.\eeq 

 \section{Construction of the remaining $N$-1 mutually unbiased bases}
 In the previous section we derived a set of two bases,
  the computational basis and the dual basis, that are mutually unbiased.
   In this section, we shall generalize this derivation in order to obtain 
   $N$-1 other mutually unbiased bases (between each other, and 
   also relatively to the computational and dual bases). 
  
  Let us denote $V^j_i$ the compositions of the shifts in the computational and 
  the dual basis:
   
 \beq
 V^j_i= V^j_{0}. V^0_{i}=\sum_{k=0}^{N-1} \gamma_{G}^{(( k\oplus_{G} i)\odot_{G} j)}\ket{ k\oplus_{G}
  i}\bra{  k}\label{defV0};i,j:0...N-1 \eeq 
 Here, the product . expresses the matricial product 
 (the usual composition law of two unitary transformations).

 As $V^0_0$ is the identity, there is no disagreement with the previous definitions.
 Note that $V^j_{0}$ and $ V^0_{j}$ do not commute: 
 \beqa V^j_{i}=V^j_{0}. V^0_{i}=
 \sum_{l=0}^{ N- l}\gamma^{l\odot_{G} j}\ket{  l}\bra{  l}.
 \sum_{k=0}^{ N- l}\ket{  k\oplus_{G} i}\bra{  k}\nonumber \\
 = \sum_{k=0}^{ N- l}\gamma^{((k\oplus_{G} i)\odot_{G} j)}
 \ket{  k\oplus_{G} i}\bra{  k}\label{global} \eeqa
 Although this expression appeared, to our knowledge, in ref.\cite{Durtmutu} for the first time, we show in the last section that the set of operators so-defined coincides with 
 the generalised Pauli group studied in ref.\cite{india}. 
 
 \beqa V^0_{i}. V^j_{0}  =
 \sum_{k=0}^{ N- l}\ket{  k\oplus_{G} i}\bra{  k}.\sum_{l=0}^{ N- l}\gamma^{l\odot_{G} j}\ket{  l}\bra{  l}\nonumber\\
 = \sum_{k=0}^{ N- l}\gamma^{(k\odot_{G} j)}\ket{  k\oplus_{G} i}\bra{  k}\nonumber\\
=\gamma^{\ominus_{G}(i\odot_{G} j)}V^j_{0}  .V^0_{i} =\gamma^{\ominus_{G}(i\odot_{G} j)}V^j_{i}\eeqa
The commutator is thus given by the following expression:
\beqa V^j_{0}. V^0_{i}-V^0_{i}. V^j_{0}  =
 (1-\gamma^{\ominus_{G}(i\odot_{G} j)})V^j_{0}. V^0_{i}\eeqa

 We recognize here a commutation rule that is known as the Weyl commutation rule,
  and was already studied a long time ago 
  \cite{weyl}. This is not astonishing because the set of unitary transformations $V^j_{i}$
   that we consider here is a discrete version of the so-called Heisenberg-Weyl group
    (compositions of translations in position and in impulsion). In dimension 2, it coincides with the Pauli group.
  When the dimension is a prime number, the field operations are the addition 
  and multiplication modulo $p$, and the properties of mutually unbiased bases are already 
  well-known in that case \cite{Ivanovic}, as well as their relation with the ``Heisenberg-Weyl-Pauli'' group 
  \cite{Englert}. 
  In the present approach, we consider,
   instead of the usual (modulo $N$)
   operations, the Galois addition and multiplication, also for  non-prime but prime power dimensions. 
   The connection with previous works related to the Pauli group approach \cite{Wootters,india,Qinfo,Zeil,Rubin}
    is established in the last section.

 By a straightforward computation, we can now derive the law of composition of these
 $N^2$ unitary transformations: 
 \beqa
 V^j_i.V^k_l=V^j_0. V^0_i.V^k_0. V^0_l\nonumber\\
 =\gamma^{\ominus_{G}(i\odot_{G} k)}V^j_0.V^k_0. V^0_i. V^0_l\nonumber\\
 = \gamma^{\ominus_{G}(i\odot_{G} k)} V^{j\oplus_G k}_{i\oplus_{G} l}
 \label{compo} \eeqa 

Up to a global phase, this looks like a groupal composition law. We shall now show that (up to phases) the $N^2$ unitary transformations 
 $V^j_{i}$ form $N+1$ commuting subgroups
of $N$ elements  that have only the identity in common. 
Moreover, each of these subgroups admits a diagonal representation in a basis that
 is mutually unbiased relatively to the the $N$ bases in which the other subgroups
  are diagonal. Note that the last property can be shown,
  following an alternative approach developed in ref. \cite{india}
  to be a consequence of the fact that the 
 $V$¥  operators, up to phases, form what is called a maximally commuting basis of
  orthogonal unitary matrices¥ (see also last section)¥. The new ingredients in our approach are (1) the expression \ref{global}
   for the generalized Pauli operators, and (2) the recognition of the fact that these operators exhibit exact groupal composition laws,
    provided (a)¥ they commute and (2)¥ they are multiplied by a convenient phase factor. 
  
In order to derive all the results, we shall take them for granted in a first time, and check afterwards that our hypothesis was correct. 
It is convenient to introduce new notations and definitions before we pursue. We shall denote
 $U^i_{l}$ the elements of these subgroups, where $i$ labels the subgroup and runs from 0 to $N$ (there are $N+1$ of them), while $l$ labels the elements of the subgroup and runs from 0 to $N-1$ (each subgroup contains $N$ elements).¥
We know already the two first subgroups, that admit a diagonal representation in the
 computational and dual bases: the first one ($i=0$) contains the elements  $V^l_0 (l:0...N-1)$, so by definition
  $¥¥U^0_{l}=V^l_0 (l:0...N-1)$. The second one contains the elements $U^1_{l}=V^0_l (l:0...N-1)$. 

  
  In virtue of the equalities \ref{transla1} and \ref{transla2}, we  can also write 
  $U^0_{l}¥=  \sum_{k=0}^{N-1} \gamma_{G}^{(k\odot_{G} l)}\ket{ k}\bra{ k}$ and $U^1_{l}¥=  
  \sum_{k=0}^{N-1} \gamma_{G}^{(k\odot_{G} l)}\ket{\tilde k}\bra{\tilde k}$. 
  A similar expression can be found for each of the $N-1$ remaining subgroups as we shall 
  now show. It is convenient at this level to parametrize the basis states that diagonalize 
  these subgroups as follows: the $k$th basis state 
  that diagonalizes the $i$th subgroup will be denoted as $\ket{  e^{i}_{k}}$. Our ultimate 
  goal is to 
  prove that there exist $N+1$ bases $\ket{  e^{i}_{k}}$ and $N^2$¥
   operators $U^i_{l}$ that are in one to one correspondence with the $V$ operators and differ from them by an appropriate phase factor, 
   such that the following constraints are fulfilled:
   \beq \label{postul}U^i_{l}¥=  \sum_{k=0}^{N-1} \gamma_{G}^{(k\odot_{G} l)}
   \ket{  e^{i}_{k}}
   \bra{  e^{i}_{k}}\ \  (l:0...N-1;i:0...N)\eeq
   \beq   \label{postul2}\braket{  e^{k}_{i}}{  e^{l}_{j}}.\braket{ e^{l}_{j}}{  e^{k}_{i}} 
   =\delta_{  k,l}\delta_{  i,j}+(1/N). (1-\delta_{ k,l})\ \  (k,l:0...N, i,j:0...N-1)\eeq
   In virtue of the commutativity of the Galois multiplication, of the identity \ref{identi2} and of the definition \ref{postul},
    the $U$ operations 
   that are labelled by a same value $i$  form a commutative subgroup and obey the 
   (exact) group composition law $U^i_l.U^i_{l'}=U^{i}_{l\oplus_G l'}$. We can guess that 
   they correspond to families of operators $V^k_{l}$ such that the (Galois) ¥ratio $k/_{G}¥l$ is constant,  because the commutation of $V^k_{l}$ and $V^{k'}_{l'}$ implies that $k'\odot_{G}l=k\odot_{G}l'$. 
   It is thus natural to try the identification $U^i_l=V_{l}^{(i-1)\odot_{G} l}$,
    up to a phase, when $i$ differs from 0 and $U^0_l=V^l_{0}$ which is consistent with our
     previous conventions. There are in general several ways to fix the phases but in any case certain constraints must be satisfied:
     
    -the phase $U^i_l/V_{l}^{(i-1)\odot_{G} l}$ is equal to 0 when $l=0$, because 
    $V_{0}^{0}=1$, and the identity is present in all subgroups
    
   -as we mentioned already, the $U$ operators must
    obey the following 
   composition law: $U^i_{l\oplus_G l'}=U^i_{l}U^i_{ l'}$, but the composition law 
   $V^j_i.V^k_l= \gamma^{\ominus_{G}(i\odot_{G} k)} V^{j\oplus_G k}_{i\oplus_G l}$
    must be guaranteed at the same time, which restricts seriously the arbitrariness in the choice of the phase.

    Let us now assume that the phase ratio between $U^i_l$ and $¥V_{l}^{(i-1)\odot_{G} l}$
     is fixed for all powers of $p$ between 0 and $m-1$ ($l=p^n, 0\leq n\leq m-1$¥)¥.
     
     We shall firstly treat the odd dimensional case. Then, 
    iterating $l$ times the composition law \ref{compo} (with $2\leq l \leq m-1$)¥, 
     we obtain the following constraints on the ratio between $U^i_{p^n\odot_{G}l}$ and 
     $¥V_{p^n\odot_{G}l}^{(i-1)\odot_{G} p^n\odot_{G}l}$, $0\leq l \leq m-1$, $0\leq n \leq m-1$:
    \beqa (U^i_{p^n\odot_{G}l})=
  \sum_{k=0}^{N-1} \gamma_{G}^{ p^n\odot_{G}k\odot_{G} l}\ket{  e_{k}^i}\bra{  e_{k}^i}=
  (U^i_{p^n})^l \nonumber\\
  = (U^{i}_{p^n}/V^{(i- 1)\odot_{G} p^n}_{p^n})^l \cdot(V^{((i- 1)\odot_{G} p^n)}_{p^n})^l   \nonumber\\
  =(U^{i}_{p^n}/V^{(i- 1)\odot_{G} p^n}_{p^n})^l 
  \cdot   \gamma_{G}^{\ominus_{G}(i-1)\odot_{G}l\odot_{G}(l\ominus_{G}1)\odot_{G}p^n\odot_{G}p^n/ _{G}2}\cdot 
  V^{((i- 1)\odot_{G} p^n\odot_{G}l)}_{p^n\odot_{G}l}\eeqa
  Here, the symbols $/ $ and $/_{G}$ indicate the multiplication by the multiplicative inverse for the usual (complex)¥ and field multiplications respectively.
   In order to fix the phase $(U^{(i\odot_{G} p^n)}_{p^n}/V^{(i-1)\odot_{G} p^n}_{p^n})$¥¥ we can make use
    of the fact that the characteristics of the field is $p$ (so to say 
    $1\oplus_G1\oplus_G...\oplus_G1$ (¥¥$p$ times¥) = 0), which implies that 
 $(U^i_{p^n})^p=1$, so that we obtain the following constraint:
 \beq(U^i_{p^n}/V^{(i-1)\odot_{G} p^n}_{p^n})^p=\gamma_{G}^{(i-1)\odot_{G}p\odot_{G}(p\ominus_{G}
 1)\odot_{G}p^n\odot_{G}p^n/ _{G}2}=1\eeq
 The phase $(U^{i}_{1}/V^{(i-1)\odot_{G} p^n}_{p^n})$¥ is thus determined up to an integer power of 
 $\gamma_{G}=e^{ i .2\pi/p}$. This is true for each integer value of $n$ between 0 and $m-1$ so that there are $p^m=N$
  different possible ways to ``fix'' the
  phases. Let us denote $\gamma_{n}¥$ the $p$th root of unity that we choose to be equal to $U^i_{p^n}/V^{(i-1)\odot_{G} p^n}_{p^n}$.  ¥¥Once this value is chosen,
   all the other phases are determined, as shows the following development:
 \beqa \label{decompo}U^i_{l}= \Pi^{m-1}_{n=0}U^i_{l_{n}\odot_{G}p^n}= \Pi^{m-1}_{n=0}(¥¥U^i_{p^n})^{l_{n}}\nonumber \\
 = \Pi^{m-1}_{n=0}\gamma_{G}^{\ominus_{G}(i-1)\odot_{G}l_{n}\odot_{G}
 (l_{n}\ominus_{G}1)
 \odot_{G}p^n\odot_{G}p^n/ _{G}2}(\gamma_{n})^{l_{n}}V^{((i-1)\odot_{G} 
 p^n\odot_{G}l_{n}¥)}_{p^n\odot_{G}l_{n}¥},\eeqa
 where the coefficients $l_{n}$¥ are unambiguously defined by the $p$-ary expansion of $l$: $l=\sum_{k=0}^{m-1}l_{n}p^n$. Moreover,
  we can check by direct computation that the $U$ operators so-defined obey an exact groupal composition law,
   independently on the choice that we could decide to perform, among the
   $p^m=N$ different possible ways to ``fix'' the
  phases $\gamma_{n}$¥:

   \beqa \label{groupodd}U^i_{l_{1}¥}.U^i_{l_{2}¥}= \Pi^{m-1}_{n=0}U^i_{l_{1n}\odot_{G}p^n}.U^i_{l_{2n}\odot_{G}p^n}
  \nonumber \\
 = \Pi^{m-1}_{n=0}\gamma_{G}^{\ominus_{G}(i-1)\odot_{G}l_{1n}\odot_{G}(l_{1n}\ominus_{G}1)\odot_{G}p^n\odot_{G}p^n/ _{G}2}
\gamma_{G}^{\ominus_{G}(i-1)\odot_{G}l_{2n}\odot_{G}(l_{2n}\ominus_{G}1)\odot_{G}p^n\odot_{G}p^n/ _{G}2}
  \nonumber \\ (\gamma_{n})^{l_{1n}}
 (\gamma_{n})^{l_{2n}} V^{((i-1)\odot_{G} p^n\odot_{G}l_{1n}¥)}_{p^n\odot_{G}l_{1n}¥}
 V^{((i-1)\odot_{G} p^n\odot_{G}l_{2n}¥)}_{p^n\odot_{G}l_{2n}¥}\nonumber \\
  = \Pi^{m-1}_{n=0}\gamma_{G}^{\ominus_{G}(i-1)\odot_{G}p^n\odot_{G}p^n\odot_{G}(l_{1n}\odot_{G}(l_{1n}\ominus_{G}1)
  \oplus_{G}l_{2n}\odot_{G}(l_{2n}\ominus_{G}1)\oplus_{G}2\odot_{G}l_{1n}\odot_{G}l_{2n})/ _{G}2}
  \nonumber \\(\gamma_{n})^{l_{1n}+_{mod p}l_{2n}}
  V^{((i-1)\odot_{G} p^n\odot_{G}(l_{1n}+_{mod p}l_{2n}))}_{p^n\odot_{G}(l_{1n}+_{mod p}l_{2n})}\nonumber \\
  = \Pi^{m-1}_{n=0}\gamma_{G}^{\ominus_{G}(i-1)\odot_{G}p^n\odot_{G}p^n\odot_{G}((l_{1n}+_{mod p}l_{2n})\odot_{G}((l_{1n}+_{mod p}l_{2n})\ominus_{G}1)/ _{G}2}
  \nonumber \\(\gamma_{n})^{l_{1n}+_{mod p}l_{2n}}
  V^{((i-1)\odot_{G} p^n\odot_{G}(l_{1n}+_{mod p}l_{2n}))}_{p^n\odot_{G}(l_{1n}+_{mod p}l_{2n})}
  \nonumber \\ = \Pi^{m-1}_{n=0}(¥¥U^i_{p^n})^{l_{1n}+_{mod p}l_{2n}}= 
  \Pi^{m-1}_{n=0}(¥¥U^i_{(l_{1n}+_{mod p}l_{2n})\odot_{G}p^n})=U^i_{l_{1}\oplus_{G}l_{2}}\eeqa

 In even prime power dimensions, the treatment is similar, although we may not divide by 2 in this case. 
 Combining the constraints $(¥¥V^{((j- 1)\odot_{G} 2^n¥)}_{2^n¥})^2=
(\gamma_{G}^{\ominus_{G}(j-1)\odot_{G}2^n\odot_{G}2^n}V^0_{0}¥)=
\gamma_{G}^{(j-1)\odot_{G}2^n\odot_{G}2^n}$ and $ U^j_{2^n}U^j_{2^n}
=U^j_{2^n\oplus_{G}2^n}=U^j_{0}=1$, we obtain the following decomposition law for the $U$ operators, which expresses their factorisation in terms of qubit operators:
 \beqa \label{decompoqubit}U^j_{l}= \Pi^{m-1}_{n=0}U^j_{l_{n}\odot_{G}2^n}=
  \Pi^{m-1}_{n=0}(¥¥U^j_{2^n})^{l_{n}}\nonumber \\
 = \Pi^{m-1}_{n=0}
 (\gamma_{G}^{(j-1)\odot_{G}2^n\odot_{G}2^n})^{{l_{n}\over 2}}
 V^{((j-1)\odot_{G} 2^n\odot_{G}l_{n}¥)}_{2^n\odot_{G}l_{n}¥},\eeqa
 where the coefficients $l_{n}$¥ are unambiguously defined by the binary expansion of $l$: $l=\sum_{k=0}^{m-1}l_{n}2^n$; $l_{n}=0¥$ or $l_{n}=1¥$.¥
 Similary to what happens in the odd dimensional case, the phase factors ¥¥¥
 $ (\gamma_{G}^{(j-1)\odot_{G}2^n\odot_{G}2^n})^{{1\over 2}}$ can be fixed with some arbitrariness,
  actually up to a sign in this case.
 Let us now check that the $U$ operators so defined obey an exact group composition law.¥
\beqa \label{groupcompoqubit}U^j_{l_{1}¥}.U^j_{l_{2}¥}= \Pi^{m-1}_{n=0}
(\gamma_{G}^{(j-1)\odot_{G}2^n\odot_{G}2^n})^{{l_{1n}+l_{2n}\over 2}}
 (¥¥V^{((j- 1)\odot_{G} 2^n¥)}_{2^n¥})^{l_{1n}+l_{2n}}=\nonumber \\
 \Pi^{m-1}_{n=0}
(\gamma_{G}^{(j-1)\odot_{G}2^n\odot_{G}2^n})^{{l_{1n}+_{mod.2}¥l_{2n}\over 2}}
(¥¥V^{((j- 1)\odot_{G} 2^n¥)}_{2^n¥})^{l_{1n}+_{mod.2}l_{2n}}=
U^j_{l_{1}\oplus_{G} l_{2}¥}.
 \eeqa
We made use of the fact that $(¥¥V^{((j- 1)\odot_{G} 2^n¥)}_{2^n¥})^2=
\gamma_{G}^{(j-1)\odot_{G}2^n\odot_{G}2^n}$, and $\gamma_{G}=-1.$¥¥

 Let us now derive an explicit expression for the $N$¥phases $¥¥U^j_{1}/V^{(j-1)\odot_{G} l}_{l}$. We shall treat separately even and odd prime power dimensions.

\subsection{Odd prime power dimensions} 
 In odd prime power dimensions, all possible consistent choices for determining the 
 phases (there are $p^m=N$ such choices¥)¥ can be expressed as follows:
 \beq \label{UsurVk}U^{i}_{1}/V^{(i-1)\odot_{G}l }_{l}=(\gamma_{G}^{\ominus(
  (i-1)\odot_{G} l\odot_{G} l)/_G 2})\gamma_{G}¥^{k\odot l},\eeq where $k$ is an arbitrary element of the field. 
  Each choice for $k$¥ (there are $N$¥  of them) leads to another 
  consistent determination of the phase ratio between the $U$¥ and $V$¥ operators. This is due to the fact that if 
  $\gamma_{G}¥^{k\odot p^n}=\gamma_{G}¥^{k'\odot p^n}, \forall n: 0\leq n\leq m-1$¥, then $k=k'$¥ in virtue of the identity \ref{identi1}.
  
 Note that, when $k=0$, which is the simplest determination of the phases, we obtain the following relation:
  \beqa \label{UsurV}
 U^i_l= (\gamma_{G}¥^{\ominus(
  (i-1)\odot_{G} l\odot_{G} l)/_{G}2})V_{l}^{(i-1)\odot_{G} l}\eeqa 
This corresponds to the choice of phases $\gamma_{n}¥=\gamma_{G}^{\ominus(
  (i-1)\odot_{G} p^n\odot_{G} p^n)/_G 2}$¥.  
It is worth noting that the relation \ref{UsurV} is also valid for $i=1$, which corresponds to the dual basis derived in the previous section.

For $i=0$, $U^{0}_{l}=V^{l}_{0}$, in agreement with the previous definitions.

  In order to check the consistency of the expression (Eqn.\ref{UsurVk}), it is sufficient, ¥
  making use of the composition law for the $V$ operators (Eqn.\ref{compo}),  to check by direct computation that the 
   $U$ operators obey an exact (so to say not up to a phase) group composition law. ¥ Remark that if
 $\tilde U^i_l=\gamma_{G}¥^{k\odot l}U^i_l$¥¥, and that $U^i_l.U^i_{l'}=U_{l\oplus_G l'}^i$, then
   $\tilde U^i_l.\tilde U^i_{l'}=\tilde U_{l\oplus_G l'}^i$. Therefore, it is sufficient to establish the groupal composition law when the expression
    (Eqn.\ref{UsurV}) is valid, in order to establish it when the expression
    (Eqn.\ref{UsurVk}) is valid, for any value of $k$.

  
  \beqa
 U^i_l.U^i_{l'}= (\gamma_{G}¥^{\ominus(
  (i-1)\odot_{G} l\odot_{G} l)/_{G}2})(\gamma_{G}¥^{\ominus(
  (i-1)\odot_{G} l'\odot_{G} l')/_{G}2})V_{l}^{(i-1)\odot_{G} l}V_{l'}^{(i-1)\odot_{G} l'}\nonumber\\
  = (\gamma_{G}^{\ominus(
  (i-1)\odot_{G} l\odot_{G} l)/_{G}2})(\gamma_{G}¥^{\ominus(
  (i-1)\odot_{G} l'\odot_{G} l')/_{G}2})(\gamma^{(\ominus_{G}(i-1)\odot_{G} 
  l\odot_{G}l')})V_{(l\oplus_G l')}^{(i-1)\odot_{G} (l\oplus_G l')}\nonumber\\
  = (\gamma_{G}¥^{\ominus(
  (i-1)\odot_{G} (l\oplus_G l')\odot_{G} (l\oplus_G l')¥)/_{G}2})V_{(l\oplus_G l')}^
  {(i-1)\odot_{G} (l\oplus_G l')}=U_{l\oplus_G l'}^i\label{compoU}; i:1...N, l,l':0...N-1\eeqa 
 
  Now that we have at our disposal an exact expression for the operators $U$,
 we can also derive an explicit expression for the $N-1$ dual bases associated to 
 the subgroups that correspond to the operators $U^i_l; i-1=1...N-1$. This can be
  realised thanks to the following identity, a direct consequence of Eqns.\ref{postul} and \ref{identi1} :
\beq \ket{  e_{k}^i}\bra{  e_{k}^i}={1\over N}\sum_{l=0}^{N-1}\gamma^{ \ominus_{G}k\odot_{G} l}U^{i}_{l}
\eeq
Obviously, if we choose another determination of the phases, so to say if we replace $U^{i}_{l}$ by $\tilde U^{i}_{l}=\gamma_{G}¥^{k'\odot l}U^{i}_{l}$, we obtain the same basis states, with their labels 
shifted by $k'$. It is thus more convenient to choose in what follows the simplest phase determination \ref{UsurV}.

By a straightforward but lengthy computation that we do not reproduce here, we obtain then the
 expression, in the computational basis, of the states of $N-1$ bases that correspond to the non-null values of the label
  $i-1$.

\beq 
  \ket{  e_{k}^i}={ 1\over \sqrt N}\sum_{q=0}^{  N-1}\gamma_{G}^
  { \ominus_{G} q\odot_{G} k}
  (\gamma_{G}¥^{(
  (i-1)\odot_{G} q\odot_{G} q)/_{G}2}) \ket{  e_{q}^0}
  \label{xxx},\eeq
  
Remark that the previous expression is also valid when $i=1$, which corresponds to 
the dual basis $\ket{\tilde j}$ (\ref{dual}).

 Let us now check by direct computation that the $N$ bases ($N-1$¥ plus the dual basis)¥ 
obtained so are
 orthonormal and mutually unbiased between each other (it is easy to check that the computational basis also fulfills these requirements)¥. Before we do so, we shall 
 rewrite the factors $¥¥(\gamma_{G}¥^{\ominus(
  (i-1)\odot_{G} q\odot_{G} q)/_{G}2})$ as follows: 
  \beq \label{conven}(\gamma_{G}¥^{\ominus(
  (i-1)\odot_{G} l\odot_{G} l)/_{G}2}) =U^{i}_{1}/V^{(i-1)\odot_{G}l }_{l}=(\gamma_{G}^{\ominus(
  (i-1)\odot_{G} l\odot_{G} l)})^{1\over 2}\eeq 
  This redefinition is less precise than the previous one, because there exist two determinations of the square root of a complex number, nevertheless we adopt it, having in mind that in odd prime power dimensions, the previous expression fixes the sign of 
  the square root of $(\gamma_{G}^{\ominus((i-1)\odot_{G} l\odot_{G} l)})$ without ambiguity. 
  We shall show in the next section that a similar expression is also valid in the even dimensional case. Let us now prove that the expression \ref{postul2} is valid.
  \beq  \braket{ e^{l}_{j} }{  e^{k}_{i}}= 
  { 1\over N}\sum_{q=0}^{  N-1}\gamma_{G}^{ \ominus_{G} q\odot_{G} (i\ominus_{G}j)}
  (\gamma_{G}^{  ((k-1)\ominus_{G}(l-1))\odot_{G} q\odot_{G}q })^{1\over 2} 
  \nonumber \eeq
  \beqa  \braket{  e^{l}_{j}}{  e^{k}_{i}}.\braket{  e^{k}_{i}}{  e^{l}_{j}}. = \nonumber \\
    { 1\over N^2}(\sum_{q=0}^{  N-1}\gamma_{G}^{ \ominus_{G} q\odot_{G} (i\ominus_{G}j)}
  (\gamma_{G}^{  ((k-1)\ominus_{G}(l-1))\odot_{G} q\odot_{G}q })^{1\over 2}).
  (\sum_{q'=0}^{  N-1}\gamma_{G}^{ \ominus_{G} q'\odot_{G} (j\ominus_{G}i)}
  (\gamma_{G}^{  ((l-1)\ominus_{G}(k-1))\odot_{G} q'\odot_{G}q' })^{1\over 2}) \nonumber \\
  = { 1\over N^2}(\sum_{q=0}^{  N-1}\gamma_{G}^{ \ominus_{G} q\odot_{G} (i\ominus_{G}j)}
  (\gamma_{G}^{  ((k-1)\ominus_{G}(l-1))\odot_{G} q\odot_{G}q })^{1\over 2}).
  (\sum_{t=0}^{  N-1}\gamma_{G}^{ \ominus_{G} (q\oplus_{G}t)
  \odot_{G} (j\ominus_{G}i)}
  (\gamma_{G}^{  ((l-1)\ominus_{G}(k-1))\odot_{G}  (q\oplus_{G}t)\odot_{G} (q\oplus_{G}t) })^{1\over 2}) \nonumber \\
  = { 1\over N^2}(\sum_{q,t=0}^{  N-1}\gamma_{G}^{ t\odot_{G} (j\ominus_{G}i)}
  (\gamma_{G}^{ 2. ((l-1)\ominus_{G}(k-1))\odot_{G} q\odot_{G}t })^{1\over 2}.
  (\gamma_{G}^{  ((l-1)\ominus_{G}(k-1))\odot_{G}  t\odot_{G} t })^{1\over 2}) 
  \nonumber \\
  = { 1\over N^2}(\sum_{q,t=0}^{  N-1}\gamma_{G}^{ ((t\odot_{G} (j\ominus_{G}i)\oplus_{G}  ((l-1)\ominus_{G}(k-1))
  \odot_{G} q\odot_{G}t) }.
  (\gamma_{G}^{  ((l-1)\ominus_{G}(k-1))\odot_{G}  t\odot_{G} t })^{1\over 2}) \nonumber \\
  = { 1\over N^2}(\sum_{t=0}^{  N-1}(\sum_{q=0}^{  N-1}
  \gamma_{G}^{ ( ((l-1)\ominus_{G}(k-1))\odot_{G}t\odot_{G} q) }).\gamma_{G}^{ (t\odot_{G} (j\ominus_{G}i)) }
  .(\gamma_{G}^{  ((l-1)\ominus_{G}(k-1))\odot_{G}  t\odot_{G} t })^{1\over 2})\nonumber \\
  = { 1\over N}\sum_{t=0}^{  N-1}\delta_{((l-1)\ominus_{G}(k-1))\odot_{G}t,0)}
.\gamma_{G}^{ (t\odot_{G} (j\ominus_{G}i)) }
  .(\gamma_{G}^{  ((l-1)\ominus_{G}(k-1))\odot_{G}  t\odot_{G} t })^{1\over 2} \nonumber \\
 =\delta_{  (l-1)\ominus_{G}(k-1),0}\delta_{  i,j}+
 (1-\delta_{ (l-1)\ominus_{G}(k-1),0}){  1\over N}\sum_{  t=0}^{  N-1}
 \delta_{  t,0}.\gamma_{G}^{ (t\odot_{G} (j\ominus_{G}i)) }
  .(\gamma_{G}^{  ((l-1)\ominus_{G}(k-1))\odot_{G}  t\odot_{G} t })^{1\over 2}\nonumber \\
  =\delta_{  k,l}\delta_{  i,j}+(1/N). (1-\delta_{ k,l})\eeqa
We made use of the fact that there is no divider of 0 excepted 0 itself 
(the multiplication $\odot_{G}$ forms a division ring). Henceforth, the following identity is valid: 
$\delta_{  a\odot_{G} b,0}=\delta_{  a,0}+(1-\delta_{  a,0})
\cdot\delta_{  b,0}$¥.

Finally, let us control the validity of the postulated expression \ref{postul}:

\beqa \label{check}  \sum_{k=0}^{N-1} \gamma_{G}^{(k\odot_{G} l)}
   \ket{  e^{i}_{k}}\bra{  e^{i}_{k}}\nonumber \\=\sum_{k=0}^{N-1} \gamma_{G}^{(l\odot_{G} k)}
   { 1\over N}\sum_{q=0}^{  N-1}\gamma_{G}^
  {  \ominus_{G}q\odot_{G} k}
  (\gamma_{G}^{  ((i-1)\odot_{G}q\odot_{G} q )})^{1\over 2}\ket{  e_{q}^0}
  \sum_{q'=0}^{  N-1}\gamma_{G}^
  {  \oplus_{G}q'\odot_{G} k}
  (\gamma_{G}^{  (\ominus(i-1)\odot_{G}q'\odot_{G} q' })^{1\over 2}\bra{  e_{q'}^0}\nonumber \\
  = { N\over N}\sum_{q,q'=0}^{  N-1}\delta_{q,q'\oplus_{G}l}(\gamma_{G}^{ ((i-1)\odot_{G}q\odot_{G} q
  \ominus(i-1)\odot_{G}q'\odot_{G} q' )})^{1\over 2}\ket{  e_{q}^0}\bra{  e_{q'}^0}
 \nonumber \\= \sum_{q'=0}^{  N-1}(\gamma_{G}^{  ((i-1)\odot_{G}(q'\oplus_{G}l)\odot_{G} (q'\oplus_{G}l)
  \ominus(i-1)\odot_{G}q'\odot_{G} q' )})^{1\over 2}\ket{  e_{q'\oplus_{G}l }^0}\bra{  e_{q'}^0}\nonumber \\
  = (\gamma_{G}¥^{(
  (i-1)\odot_{G} l\odot_{G} l)})^{1 \over 2}\sum_{q'=0}^{  N-1}\gamma_{G}^{  (((i-1)\odot_{G}l)\odot_{G}q')}
  \ket{  e_{q'\oplus_{G}l}^0}\bra{ e_{q'}^0}=U^i_{l}(l:0...N-1;i:1...N)\eeqa

   \subsection{Even prime power dimensions}
   In this case the explicit expressions for the mutually unbiased bases are less easy to manipulate. Once again, there are $p^m$ ($2^m$ in this case) possible ways to determine the 
   phases $¥¥U^j_{1}/V^{(j-1)\odot_{G} l}_{l}$, but they are equivalent, up to a relabelling of the basis states.
   
   In the next development, we shall implicitly choose a certain determination of the 
   square root of $\gamma_{G}^{(j-1)\odot_{G}2^n\odot_{G}2^n}$¥¥
    that is equal to $i^{(j-1)\odot_{G}2^n\odot_{G}2^n}$¥.  
   \beqa U^j_{l}= \Pi^{m-1}_{n=0}U^j_{l_{n}\odot_{G}2^n}= \Pi^{m-1}_{n=0}(¥¥U^j_{2^n})^{l_{n}}\nonumber \\
 = \Pi^{m-1}_{n=0,l_{n}\not= 0}(\gamma_{G}^{(j-1)\odot_{G}2^n\odot_{G}2^n})^{{1 \over 2}}
 (¥V^{((j-1)\odot_{G} 2^n¥)}_{2^n¥})^{l_{n}}\nonumber \\
 = \Pi^{m-1}_{n=0, l_{n}\not= 0¥}i^{(j-1)\odot_{G}2^n\odot_{G}2^n}
 (¥V^{((j-1)\odot_{G} 2^n¥)}_{2^n¥})\nonumber \\
 = ¥¥(¥¥\Pi^{m-1}_{n=0, l_{n}\not= 0¥}i^{(j-1)\odot_{G}2^n\odot_{G}2^n}\gamma_{G}^{(j-1)\odot_{G}2^n\odot_{G}2^{n'}})
 V^{((j-1)\odot_{G} l¥)}_{l¥}
 \label{UsurVeven}\eeqa
 where the coefficients $l_{n}$¥ are unambiguously defined by the $p$-ary (here binary) ¥expansion of $l$: $l=\sum_{k=0}^{m-1}l_{n}2^n$, while 
 $n' $ is the smallest integer strictly larger than $n$ such that $l_{n'}\not= 0$, if it exists, 0 otherwise. 
 
 This result is a generalisation of the identity \ref{UsurV}, because in both cases the phases are square roots of integer powers of gamma:
 
 $¥¥(¥U^j_{1}/V^{(j-1)\odot_{G} l}_{l})^2=\gamma_{G}¥^{\ominus(
  (j-1)\odot_{G} l\odot_{G} l)}$.
  
  Nevertheless, in the present case we obtain the following determination of the square root
   of $(\gamma_{G}¥^{\ominus(
  (j-1)\odot_{G} l\odot_{G} l)})$: 
  ¥¥\beqa \label{explicit}
   (\gamma_{G}¥^{\ominus( (j-1)\odot_{G} l\odot_{G} l)})^{{1 \over 2}}=
   (\gamma_{G}¥^{\oplus( (j-1)\odot_{G} l\odot_{G} l)})^{{1 \over 2}}\nonumber \\
   = ¥¥\Pi^{m-1}_{n=0, l_{n}\not= 0¥}i^{(j-1)\odot_{G}2^n\odot_{G}2^n}
  \gamma_{G}^{(j-1)\odot_{G}2^n\odot_{G}2^{n'}}\eeqa
  where $n'$ and $l_{n}¥$¥ were defined previously. What is particular with even prime powers is that the square root of an integer power 
  of $\gamma_{G}$¥ is not an integer power of $\gamma_{G}$ as was the case in odd prime power dimensions.¥ We are forced 
  to introduce +i and -i. What is interesting is that this minimal 
  extension is sufficient in order to diagonalize the operators of the
   generalized Pauli group in even prime power dimensions, a fact that was already
    recognized in previous references on the subject (\cite{Wootters,rotteler})¥.

    Combining the equations \ref{xxx} and \ref{conven}, we obtain the following synthetic expression:
    
    \beqa 
  \ket{  e_{k}^i}={ 1\over \sqrt N}\sum_{q=0}^{  N-1}\gamma_{G}^
  { \ominus_{G} q\odot_{G} k}
  (\gamma_{G}¥^{(
  (i-1)\odot_{G} q\odot_{G} q)})^{{1 \over 2}} \ket{  e_{q}^0}
  \label{synthetic}\eeqa
  Taking account of the Eqn.\ref{explicit} we get an explicit expression for the mutually unbiased bases:
  \beqa
  \ket{  e_{k}^i}={ 1\over \sqrt N}\sum_{q=0}^{  N-1}\gamma_{G}^
  { \ominus_{G} q\odot_{G} k}
 \Pi^{m-1}_{n=0, q_{n}\not= 0¥}i^{(j-1)\odot_{G}2^n\odot_{G}2^n}
  \gamma_{G}^{(j-1)\odot_{G}2^n\odot_{G}2^{n'}}\ket{  e_{q}^0}
  \label{yyy},\eeqa
where $q=\sum_{k=0}^{m-1}q_{n}2^n$, while 
 $n' $ is the smallest integer strictly larger than $n$ such that $q_{n'}\not= 0$, if it exists, 0 otherwise.

As a consequence of the composition laws \ref{compo} and \ref{groupcompoqubit},
 \beqa U^j_{l_{1}}.U^j_{l_{2}}=
 (\gamma_{G}^{(j-1)\odot_{G}l_{1}\odot_{G}l_{1}})^{{1\over 2}}
 .(\gamma_{G}^{(j-1)\odot_{G}l_{2}\odot_{G}l_{2}})^{{1\over 2}}
 \gamma^{(j-1)\odot_{G}(l_{1}\odot_{G} l_{2})} V^{(j- 1)
 \odot_{G}(¥¥l_{1}\oplus_{G}l_{2})}_{l_{1}\oplus_{G}l_{2}}¥\nonumber \\
 =(\gamma_{G}^{(j-1)\odot_{G}(¥¥l_{1}\oplus_{G}l_{2})\odot_{G}(¥¥l_{1}\oplus_{G}l_{2})})^{{1\over 2}}
 .V^{(j- 1)
 \odot_{G}(¥¥l_{1}\oplus_{G}l_{2})}_{l_{1}\oplus_{G}l_{2}}=U^j_{l_{1}\oplus_{G}l_{2}¥}\eeqa¥
Therefore,\beq(\gamma_{G}^{(j-1)\odot_{G}l_{1}\odot_{G}l_{1}})^{{1\over 2}}
 .(\gamma_{G}^{(j-1)\odot_{G}l_{2}\odot_{G}l_{2}})^{{1\over 2}}
 .\gamma^{(j-1)\odot_{G}(l_{1}\odot_{G} l_{2})}=(\gamma_{G}^{(j-1)\odot_{G}(l_{1}\oplus_{G}l_{2})
 \odot_{G}(l_{1}\oplus_{G}l_{2})})^{{1\over 2}}.\label{simili}\eeq
 Formally we can rewrite the previous equation as follows:
 $ (\gamma_{G}^{(j-1)\odot_{G}(¥¥a\oplus_{G}b)\odot_{G}(¥¥a\oplus_{G}b)})^{{1\over 2}} =
(\gamma_{G}^{(j-1)\odot_{G}(¥¥a\odot_{G}a)})^{{1\over 2}}.(\gamma_{G}^
{(j-1)\odot_{G}(¥¥b\odot_{G}b)})^{{1\over 2}}.(\gamma_{G}^{2.((j-1)\odot_{G}a
\odot_{G}b)})^{{1\over 2}}, $
 which is reminiscent of the equation \ref{identi2}, although we are dealing here with 
half integer powers of $\gamma_{G}$ instead of integer powers.
 Thanks to this property, it is possible to reproduce nearly litterally the proofs given
  in odd prime power dimensions of the validity of the identities \ref{postul}
   and \ref{postul2}, because the automatisms of computation are nearly equivalent. 
   It is important to note however that in even prime power dimensions the 
   expressions of the type $(\gamma_{G}^{(¥¥a\odot_{G}a)})^{{1\over 2}}$ do well
    represent square roots of $\gamma_{G}^{(¥¥a\odot_{G}a)}$¥ but must 
   be considered as functions that depend on the $2^m$¥ variables $a$ instead of only two variables, as would
    be the case if we considered litterally square roots of integer powers of 
   $\gamma_{G}$ (with $\gamma_{G}=-1$)¥. When $a$¥ is specified, the sign of the square root is also specified,
    according to the explicit expression \ref{explicit}. The even and odd dimensional cases are covered by the synthetic expression
     \ref{synthetic}.

\section{Open questions, comments and conclusions.}
\subsection{Other symmetries}

At first sight, the computational basis plays a special 
role in our approach, but one can show that, to some extent,
 all the mutually unbiased bases can be treated on the same footing. 
 This can be seen as follows. Now that we have at our disposal an explicit expression (Eqns.\ref{xxx},\ref{yyy}) 
 for all the mutually unbiased bases,
   we can ``reevaluate the situation from the point of view of one of them'', say the $i$th basis (with $i$ different from zero)¥.
    In order to do so, we can express the action of the operator 
  $V^m_{n}$ in terms of its basis states. After a straightforward computation, we get that 
  $V^n_{m}(0)¥=phase.V^{m}_{\ominus_{G}n\oplus_{G}(i-1)\odot_{G}m}(i)¥$, where
   $V^n_{m}(0)¥=
  \sum_{k=0}^{N-1} \gamma_{G}^{(( k\oplus_{G} m)\odot_{G} n)}
  \ket{ e_{k\oplus_{G} m}^0}\bra{  e_{k}^0}$ and $V^n_{m}(i)¥=
  \sum_{k=0}^{N-1} \gamma_{G}^{(( k\oplus_{G} m)\odot_{G} n)}\ket{ e_{k\oplus_{G} m}^i}\bra{  e_{k}^i}; i:1...N$.
   These relations (that we give without proof but are easy to derive from Eqns.\ref{xxx},\ref{yyy})
   are bijective. So the whole discrete Heisenberg-Weyl group is invariant (up to permutations and phase shifts) when we reexpress it in any of the $N+1$ mutually unbiased
    bases. We shall not develop this question here, but this property has important implications in the theory of cloning machines, 
    in relation with error operators and optimal cloning (\cite{DurtNagler,qutrits}). In prime dimensions, the invariance of the Heisenberg-Weyl group under conjugation by any unitary matrix that maps the computational basis onto a mutually unbiased basis is a basic property of a larger group that is known as the Jacobi or Clifford group and 
    possesses many applications in number theory and quantum computing \cite{Grassl}.
    
    Beside, there exists a one to one correspondence between generalised Bell states \cite{Durtmutu} and the Heisenberg-Weyl group. 
    The properties of invariance of the Bell states in mutually unbiased bases
     appeared to be very useful in the resolution of the so-called mean king problem 
  \cite{Englert,vaid,2003}, where it also led to a compact
   and elegant expression valid in all prime power dimensions \cite{Durtmean}.

\subsection{Connection with previous works.}

The expression
 of the states of the mutually unbiased bases that we derived in the present paper 
 ($\ket{  e_{k}^i}={ 1\over \sqrt N}\sum_{q=0}^{  N-1}\gamma_{G}^
  { \ominus_{G} q\odot_{G} k}
  (\gamma_{G}^{  ((i-1)\odot_{G}q\odot_{G} q )})^{1\over 2}\ket{  e_{q}^0}$¥) is actually equivalent to the solution derived by Ivanovic \cite{Ivanovic} when the dimension is an 
 odd prime (which can be shown, when rewritten according to our conventions, to be equivalent to the expression 
 $\ket{  e_{k}^i}^{  Ivan.}={ 1\over \sqrt N}\sum_{q=0}^{  N-1}\gamma_{G}^
  { \ominus_{G} q\odot_{G} k}
  (\gamma_{G}^{  ((i-1)\odot_{G}q\odot_{G} q )})\ket{  e_{q}^0}$¥).
   Our expression differs by a factor $1/_{G}¥2$¥ in one of the exponents of $\gamma_{G}$. 
  When the dimension is prime and odd, it is easy to compensate the difference 
  by a relabelling of the basis states, because the division by 2 is a permutation of the finite fields with $p$ elements when $p$ is a prime odd number, 
  but contrary to Ivanovic's expression, our expression is easily generalized in even prime dimension 2
   (the qubit case), in which
   case we rederive the eigen bases of the Pauli operators, and in prime power dimensions. 
   
  As it was shown by Wootters and Fields \cite{Wootters}, the generalisation in prime power dimensions of Ivanovic's expression is the following:
  
  \beq¥¥\ket{  e_{k}^i}={ 1\over \sqrt N}\sum_{q=0}^{  N-1}\gamma_{G}^
  {Tr.( \ominus_{G} q\odot_{G} k)}
  (\gamma_{G}^{  Tr.(r\odot_{G}q\odot_{G} q )})\ket{  e_{q}^0},\label{USA}\eeq where $Tr.$¥ represents the field theoretical trace.
    Our expression seems to be a bit simpler, but both expressions require to know the addition and multiplication tables of the field, so that the apparent gain in simplicity is relative. 
    Moreover, both expressions are equivalent up to a relabelling in odd prime power dimensions as we shall now show. We could
     establish the equivalence at once but we prefer to base our derivation on the results of the reference 
    \cite{india} where the interrelation between the Pauli group approach and the expression of Wootters and Fields with the trace factor is made 
    (section 4.3., \cite{india}).
   Actually, there exist also different groups that present properties similar to those of the generalized Pauli group
    \cite{Zeil} but do not obey the definition that we gave here.  
    Nevertheless, the generators of the two subgroups corresponding to 
    the computational and the dual basis that are given in the reference \cite{india} coincide with our choice (the subgroups $V^l_{0}¥$¥ and $V^0_{l}¥$¥
     correspond to the classes $C_{0}¥$¥ and $C_{1}¥$
     studied in ¥\cite{india})¥ so that we are talking exactly about the same group, up to phases. As, in
     the same paper, a relation was established with the solution of Wootters and Fields \cite{Wootters}, 
     our expression for mutually unbiased bases must necessarily coincide with the expression derived by Wootters and Fields.
      In the reference \cite{india}, 
     it is shown that when there exists a maximal commuting basis of orthogonal unitary matrices, 
     the $N+1$ bases that diagonalize these classes are unabiguously defined and, moreover, are mutually unbiased. A maximal commuting basis of
      orthogonal unitary matrices is a set of $N+1$¥ sets of $N-1$ commuting unitary operators (or classes) 
      plus the identity such that these $N^2$¥ operators are orthogonal regarding the in-product
       induced by the (usual operator)¥ trace denoted $tr.$¥. 
      It is easy to show that the $V$ operators defined in Eqn.\ref{defV0}
       are unitary with $(¥¥V^j_i)^+=(¥¥V^j_i)^{-1}=\gamma^{\ominus_{G}(i\odot_{G} j)}
       V^{\ominus_G i}_{\ominus_G j}$ and that $tr.V^j_i=N.\delta_{i,0}.\delta_{j,0}¥$¥. Making 
       use of the composition law  \ref{compo}, we obtain the relation $tr.((¥¥V^j_i)^+.V^k_l)
       =N.\delta_{i,l}.\delta_{j,k}¥$¥ so that they form a 
       maximal commuting basis of unitary operators. This theorem suggests another way to derive an expression for the mutually unbiased bases: it is sufficient to find the 
       common eigenstates of the classes of operators $V_{l}^{(i-1)\odot_{G} l}$ (where $l$ varies from 0 to $N-1$¥¥) in order to determine the value of the 
       states of the $i$th mutually unbiased basis. When the dimension is an odd (even) prime power¥, one can check 
       by direct substitution of the expression \ref{xxx} (\ref{yyy})¥¥ that the states 
       $\ket{  e_{k}^i}$¥ are common eigenstates of the $i$th¥ class: 
       
       \beqa
  V_{l}^{(i-1)\odot_{G} l}\ket{  e_{k}^i}=
  \sum_{k'=0}^{N-1} \gamma_{G}^{(( k'\oplus_{G} l)\odot_{G} (i-1)\odot_{G} l)}\ket{ k'\oplus_{G}
  l}\bra{  k}{ 1\over \sqrt N}\sum_{q=0}^{  N-1}\gamma_{G}^
  { \ominus_{G} q\odot_{G} k}
  (\gamma_{G}¥^{(
  (i-1)\odot_{G} q\odot_{G} q)/_{G}2}) \ket{  e_{q}^0} \nonumber\\
  =
  { 1\over \sqrt N}\sum_{q=0}^{N-1} \gamma_{G}^{(( q\oplus_{G} l)\odot_{G} (i-1)\odot_{G} l)}
  \gamma_{G}^
  { \ominus_{G} q\odot_{G} k}
  (\gamma_{G}¥^{(
  (i-1)\odot_{G} q\odot_{G} q)/_{G}2}) \ket{  e_{q\oplus_{G}l}^0}\nonumber\\
  =
  \gamma_{G}^{  (l¥\odot_{G} k)}
   \gamma_{G}^{( (i-1)\odot_{G} l\odot_{G} l)/_{G}2)}{ 1\over \sqrt N}\sum_{q\oplus_{G}l=0}^{N-1}
  \gamma_{G}^{ \ominus_{G} (q\oplus_{G}l)¥\odot_{G} k}
  (\gamma_{G}¥^{((i-1)\odot_{G} (q\oplus_{G}l)¥\odot_{G} (q\oplus_{G}l)¥)/_{G}2})
   \ket{  e_{q\oplus_{G}l}^0}\label{zzz}\eeqa
   Thanks to the product law \ref{simili}, the proof is entirely similar in even prime power dimensions.
   
   In order to establish explicitly and once for all 
   the equivalence between the expression of Wootters and Fields (¥¥\ref{USA})
    and ours (¥¥\ref{xxx}), some work remains to be done because in the reference 
    \cite{india} no proof is given of the fact that the expression 
    \ref{USA} represents eigenstates of the generalised Pauli operators. In order to prove this result, it is useful to 
   introduce two (field theoretical) ¥dual bases: the first one, which is dual relatively to the trace contains $m$ elements $\tilde p$ of the field such
    that $Tr.p^i\odot_{G}\tilde p^j=\delta_{i,j}¥$; the second one  which is dual
     relatively to the rest after division by $p$¥ contains $m$ elements $\tilde{\tilde p^j}$ of the field such
    that $(p^i\odot_{G}\tilde{\tilde p^j})_{0}¥=\delta_{i,j}¥$, where $(q)_{0}¥¥$¥ represents (we work in dimension $p^m$¥)¥ 
    the rest after division of $q$ by $p$. These bases can be shown to exist and to be unique, in virtue of the fact that the bilinear 
    forms $Tr.(¥¥x\odot_{G} y)$ and $(x\odot_{G}y)_{0}¥¥$¥¥ are non-singular 
    \cite{Karpilovsky}, a direct consequence of the identity \ref{identi1}. Beside, the $m$ generators
     of the $k$th ¥class considered in \cite{india} are equal to 
    $X^i.\Pi_{l=0}^{m-1}(¥¥\Pi_{j=0}^{m-1}(¥¥Z^j)^{b^l_{ij}¥}))^{k_{l}¥}$¥, with $j,l:0...m-1, $. In the previous expression,
     the 
    coefficients $k_{l}$¥ are unambiguously defined by the $p$-ary expansion of $k$: $k=\sum_{l=0}^{m-1}k_{l}p^l$ while the multiplication matrix $b$¥ is defined as follows: 
    $\gamma_{i}\odot_{G}\gamma_{j}¥=\sum_{l=0}^{m-1}b^l_{ij¥}\gamma_{l}¥$,
     where the $\gamma$'s are a basis of the field (here we shall consider without loss of generality that 
    $\gamma_{i}=p_{i}¥$¥)¥. The operators $X^i$ are ``local'' operators that shift
     the $i$th 
    component of the label of the $k$th ¥basis state $¥¥\ket{  e_{k}^0}$ 
    (with $k=\sum_{l=0}^{m-1}k_{l}p^l$)¥ by unity (modulo $p$¥)¥: $X^i\ket{  e_{k}^0}=\ket{  e_{k'}^0}$ with 
    $k'_{i}=k_{i}+_{mod.p}1, k'_{l}=k_{l}¥$ when $l\not= i$. The operator $Z^j$¥ multiplies the state $¥¥\ket{  e_{k}^0}$
     by 
    a global phase equal to $¥\gamma_{G}^{k_{¥¥j¥}}$¥¥¥¥¥. In our approach, the generators of the $k'$th class can be shown to be the same, provided 
    the coefficients of $k'$ expressed in the double-tilded dual basis $\tilde{\tilde p^j}$ 
    defined here above are the same as those of $k$¥¥ in the direct basis $p$¥: $k=\sum_{l=0}^{m-1}k_{l}p^l$ and 
    $k'=\sum_{j=0}^{m-1}k_{j}\tilde{\tilde p^j}$. 
     Beside, the expressions with and without Trace \ref{xxx}  and \ref{USA}
      are equivalent, up to a bijective relabelling, in virtue of the following identity:
     $Tr.(r\odot_{G}k )=((r'/_{G}¥ 2)\odot_{G}k )_{0}¥$ with $k$ and $¥r$ arbitrary elements of the Galois field with
      $N=p^m$¥ elements, $r=\sum_{l=0}^{m-1}r_{l}\tilde p^l$¥ and 
      $r'¥=\sum_{l=0}^{m-1}r_{l}\tilde{\tilde p^l}\odot_{G}2$.
      
      This comparison emphasises the difference between our approach and previous approaches: our expression
       \ref{defV0} of the generalised Pauli operators is global and non-local, although they can be decomposed as products of local operators. It also shows that 
       the field theoretical trace is replaced in our approach by another non-singular bilinear form, the rest after division by $p$.¥
       
       Although it is out of the scope of the present paper, it would be interesting to understand the relation between our results in even prime power dimensions and the results presented in references \cite{Wootters,rotteler}.

     \subsection{Other dimensions.}

It is still an open question to know whether maximal sets of mutually unbiased bases exist in arbitrary 
dimensions. For instance in dimension 6 which is the smallest dimension that is not a power of a prime, nobody knows whether or not such
 a maximal set exists \cite{Archer,Grassl}. It is not possible to apply our treatment in this case because no finite field with 6 elements exists. We could try to repeat
  the procedure with operations that do not form a field; for instance we could try to find a distributive ring with 6 elements 
  (such a ring obeys the same definition as a field (the definition that was given at the beginning of the paper),
   excepted that the multiplication needs not be
 invertible-dividers of zero different from zero are allowed). One can show that there is only one distributive ring with 6 elements, that corresponds to the usual operations (multiplication and addition modulo 6). 
 If we study the structure of the $N^2=36$ Heisenberg-Weyl unitary transformations in that
 case, we find that there are more than $N+1=7$ subgroups of 6 elements (5+the identity). This is because, 
 as a consequence of the non-invertibility of the multiplication modulo 6 (3 and 2 divide zero),
  certain operators present degeneracies and belong simultaneously to different subgroups (a treatment of similar type is given in detail in the reference \cite{Durtmutu} for the case $N=4$¥ )¥¥. 
 The bases that diagonalize these operators are not mutually unbiased in general and the construction that was succesfully applied in prime power dimensions does not provide a 
 maximal set of mutually unbiased bases. Therefore the question of the existence of 7 
 mutually unbiased bases in a 6 dimensional Hilbert space is still open, and our approach does unfortunately not contribute to the elucidation of that problem.

\subsection{Conclusions}
As we already mentioned, there is a one to one correspondance between (generalised) Bell states
 and (generalised) Pauli operators \cite{Durtmutu} (see also \cite{Planat} 
 for a different approach based on additive and multiplicative characters of the Galois field).
   It can also be proven \cite{Durtmean,Durtmutu} that the Bell states are invariant
    when we pass from one of the mutually unbiased bases to another one, 
   an important result in the theory of cloning machines that was only conjectured until now
    \cite{qutrits}. Actually, the present results were largely inspired by results
    that we obtained in the framework of quantum cryptography \cite{DurtNagler} where 
    the interest of mutually unbiased has been recognised several years
     ago, for what concerns encryption \cite{BB84,optimalencrypt,Bechmann} and cloning
      as well \cite{DurtNagler,qutrits,CERFPRL}.

   It is worth noting that, beside quantum cryptography and quantum cloning, the Bell states found also many applications
    in quantum teleportation and dense coding and the connections between mutually unbiased bases, complete orthogonal families of unitary 
    matrices, and teleportation, were already emphasised in the past \cite{Fivel,werner}. There exists also an impressive litterature
     about the interrelation between 
    finite fields and discrete Wigner representation \cite{Wootters2,discretewigner,paz}. It is worth noting that if we perform
      a tomographic development of the density matrix in the basis of the $V$ operators, we obtain \cite{Durtmutu} the following identity:
      $\rho=(1/N)\sum_{k,l}V^{k}_{l}Tr.((V^{k}_{l})^+.\rho)$. 
      
      It is very instructive to compare the amplitudes of the decomposition of a density matrix into the $V$ basis
    with the Wigner function:
    
    $Tr.(V^{k}_{l})^+\rho$=$Tr.(V^{k}_{l})^+\sum_{i,j=0}^{N-1}\rho_{i\oplus j,i}
    $$\ket{  e_{i\oplus j}^0}\bra{  e_{i}^0}$=$\sum_{i,j=0}^{N-1}\rho_{i\oplus j,i}
    \gamma^{\ominus(i\oplus l)\odot k}\delta_{l,j}$=
    
    $\gamma^{\ominus l\odot k}\sum_{j=0}^{N-1}\rho_{j\oplus l,j}
    \gamma^{\ominus j\odot k}.$
    
    The Wigner function can be written as follows \cite{peres} in terms of the conjugate continuous variables $q$ and $p$¥¥:
    
    $W(q,p)=C.\int d^3r
     \rho(q-r,\ q+r)e^{2.i.p.r/h}=C'.\int d^3r'
     \rho(r',\ 2.q-r')e^{-i.2.p.r'/h}$ where $C$ is a normalisation constant, while $i$ is the square root of -1. 
     The analogy of both expressions is striking and, as we can see, the 
   tomography of a quantum state that we realize in the Pauli group approach provides a discrete counterpart
    of the Wigner representation. In general the coefficients  $Tr.((V^{k}_{l})^+.\rho)$ 
    are complex, which does not 
    meet the requirements of a properly discretized Wigner function, but in even prime
     power dimensions this difficulty can be overcomen because the $V$ operators are 
     Hermitian up to a global phase that was defined in Eqn.\ref{explicit} (the $U$ operators defined by Eqn.\ref{UsurVeven} are Hermitian and unitary and also 
     provide¥ an orthogonal basis)¥. It is out of the scope of the present paper but it would be interesting to study the connection with other proposals for discrete 
    phase-space representation¥¥ \cite{Wootters2,discretewigner,paz}. Note that as the $V$¥ operators are diagonal in the $N+1$ mutually unbiased bases, full tomography can be obtained 
      by performing $N+1$¥ von Neumann measurements, as was already shown by Ivanovic in prime dimensions and Wootters and Fields in prime power dimensions.¥
      
       Finally, the properties of Bell states are also directly related to the error operators \cite{errorcorr,nielsen,paz}, and it would be worth investigating to which extent our formalism contributes
    to a simplification of the theory of error correcting codes, in prime power dimensions.

To conclude, we note that, despite of the fact that the problem (and its solutions) seem
 to be regularly rediscovered by different generations of physicists, which means also
  a lack of time and energy, our results about the Mean King's problem \cite{Durtmean}
   confirm that it is important to explore alternative
 approaches in the treatment of the 
question of mutually unbiased bases. 

We wish that the present paper contributes to a deeper understanding of the old problem of mutually unbiased bases.

    \leftline{\large \bf Acknowledgment}
\medskip The author acknowledges
a Postdoctoral Fellowship of the Fonds voor Wetenschappelijke Onderzoek,
Vlaanderen and also support from the
IUAP programme of the Belgian government, the grant V-18, and the Solvay Institutes for Physics and Chemistry.

 Sincere thanks to profs. P.Cara (VUB), E.Jespers (VUB) and B-G Englert (NUS)
 for fruitful and enjoyable 
discussions and advice.

{\bf Appendix: Field and modulo $N$¥ operations for $N=4$¥.}
\newpage
 
 \begin{table}
\begin{tabular}{c||c|c|c|c}
\hline $\odot_{G}$  & $0$ & $1$ & $2$  & $3$\\
\hline \hline  0 & 0 & $0$ & $0$  & $0$ \\
1 & $0$ & $1$ & $2$  & $3$\\
2 & $0$ & $2$ & $3$  & $1$\\
3 & $0$ & $3$ & $1$  & $2$ \\
 \hline
\end{tabular}
\caption{The field multiplication in dimension 4. }\label{tab1}

\begin{tabular}{c||c|c|c|c}
\hline $\oplus_{G}$  & $0$ & $1$ & $2$  & $3$\\
\hline \hline  0 & 0 & $1$ & $2$  & $3$ \\
1 & $1$ & $0$ & $3$  & $2$\\
2 & $2$ & $3$ & $0$  & $1$\\
3 & $3$ & $2$ & $1$  & $0$ \\
 \hline
\end{tabular}
\caption{The field addition in dimension 4. }\label{tab2}
 \end{table}
  \begin{table}
\begin{tabular}{c||c|c|c|c}
\hline $. mod 4$  & $0$ & $1$ & $2$  & $3$\\
\hline \hline  0 & 0 & $0$ & $0$  & $0$ \\
1 & $0$ & $1$ & $2$  & $3$\\
2 & $0$ & $2$ & $0$  & $2$\\
3 & $0$ & $3$ & $2$  & $1$ \\
 \hline
\end{tabular}
\caption{The multiplication modulo 4. }\label{tab3}

\begin{tabular}{c||c|c|c|c}
\hline $+ mod 4$  & $0$ & $1$ & $2$  & $3$\\
\hline \hline  0 & 0 & $1$ & $2$  & $3$ \\
1 & $1$ & $2$ & $3$  & $0$\\
2 & $2$ & $3$ & $0$  & $1$\\
3 & $3$ & $0$ & $1$  & $2$ \\
 \hline
\end{tabular}
\caption{The addition modulo 4. }\label{tab4}
\end{table}

\end{document}